\documentclass[twocolumn, 10pt, aps, pra, superscriptaddress,groupedaddress]{revtex4}

\usepackage{graphicx}
\usepackage{amsmath}
\usepackage{amssymb}
\usepackage{bm}
\usepackage{amsfonts}
\usepackage[usenames, dvipsnames]{xcolor}
\usepackage{appendix}
\usepackage[colorlinks=true, pdfstartview=FitV, linkcolor=blue, citecolor=blue, urlcolor=blue]{hyperref}
\usepackage{color}
\usepackage{epstopdf}

\begin{document}
 
\title{Pathway to chaos through hierarchical superfluidity \\ in blue-detuned cavity-BEC systems}
 
\author{Rui Lin}
\affiliation{Institute for Theoretical Physics, ETH Zurich, 8093 Zurich, Switzerland}
\author{Paolo Molignini}
\affiliation{Institute for Theoretical Physics, ETH Zurich, 8093 Zurich, Switzerland}
\affiliation{Clarendon Laboratory, University of Oxford, Parks Road, Oxford OX1 3PU, United Kingdom}
\author{Axel U. J. Lode}
\affiliation{Institute of Physics, Albert-Ludwig University of Freiburg, Hermann-Herder-Stra{\ss}e 3, 79104 Freiburg, Germany}
\author{R. Chitra}
\affiliation{Institute for Theoretical Physics, ETH Zurich, 8093 Zurich, Switzerland}

 
\begin{abstract}
We explore the role of atomic correlations in a harmonically trapped Bose-Einstein condensate coupled to a dissipative cavity, where both the atoms and the cavity are blue detuned from the external pumping laser.  Using a genuine many-body approach that goes beyond mean-field, we extract density distributions and many-body correlations to unveil a pathway to chaos at large pump power through a hierarchical self-organization of the atoms, where the atoms transition from a single-well optical lattice to a double-well optical lattice. Correlated states of the atoms emerge and are characterized by local superfluid correlations in phases which are globally superfluid or Mott insulating. Local superfluid-Mott transitions are precluded by a dynamical instability to chaos which occurs via quasiperiodic attractors.  Our results explain the mechanism behind the dynamical instabilities observed in experiments.
\end{abstract}

\maketitle

\emph{Introduction} -
Experimental advances in the past decade have heralded a new era in light-matter hybrid systems, where  quantum light is used to engineer correlated phases of matter.
In the solid state realm, coupling to light has  been used to activate phases of matter like ferroelectricity~\cite{zalden19} and superconductivity~ \cite{schlawin19}.  In the quantum engineering domain, cavity-QED systems with their tunable light matter couplings provide a versatile platform to realize hybrid correlated quantum fluids like polaritons~\cite{knuppel19,latini19}, permit the encoding of qubits through photons~\cite{korber18}, and generate entangled quantum many-body states for quantum computation~\cite{welte18}.

A landmark example of a light induced phase is superradiance ~\cite{dicke54,hepp73,wang73,carmichael73} in strongly coupled cavity-Bose-Einstein condensates (BEC) systems where  the atoms in the  BEC  self-organize  onto a lattice dynamically generated by the cavity field ~\cite{nagy08,baumann10,baumann11}.
Cavity-BEC systems also host complex phases like Mott insulators~\cite{klinder15,bakhtiari15,flottat17,axel17,lin19}, supersolids \cite{leonard17,morales18,mivehvar18}, and spin textures ~\cite{mivehvar172,axel18,landini18,mivehvar19,chiacchio19}. They additionally allow the simulation of many-body Hamiltonians having no solid-state counterparts, like spin models with both short and long range interactions~\cite{vaidya18,schlawin19}, and the realization of exotic collective magnetic phenomena~\cite{davis19}.

\begin{figure}
	\centering
	\includegraphics[width=\columnwidth]{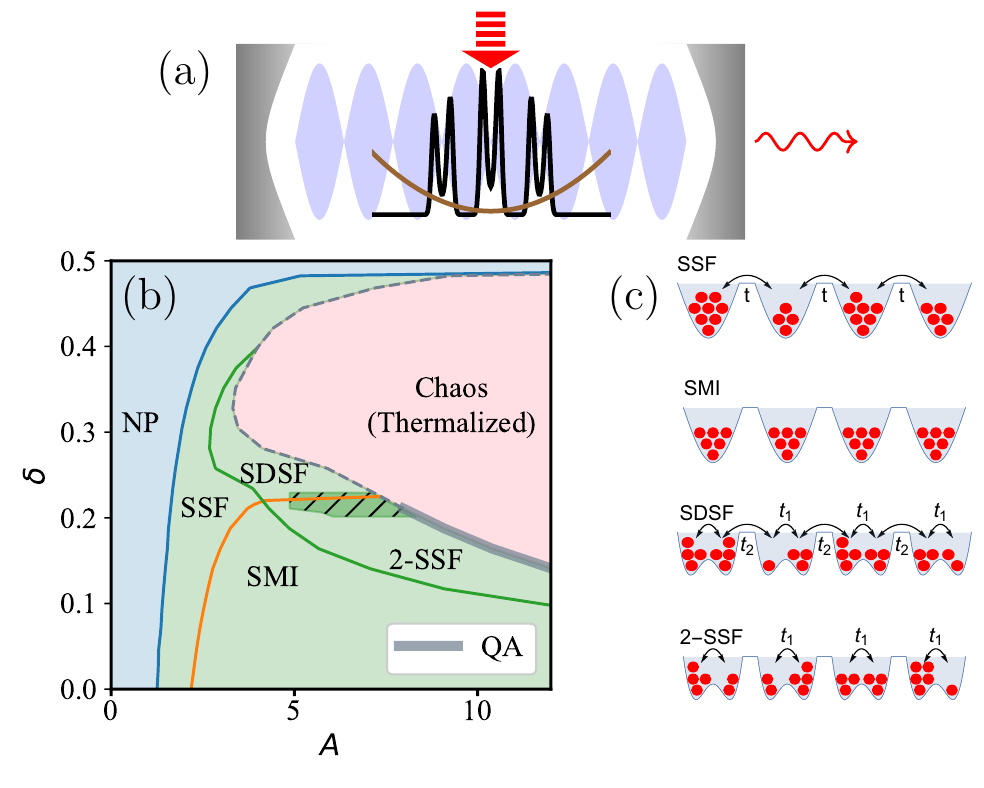}
	\caption{
	(a) Schematic setup of a trapped  cavity-BEC.
	(b) Phase diagram.  For cavity detuning $0<\delta<1/2$,  the system transitions from the normal phase to  the dynamically unstable region via  superradiance  with increasing pump rate $A\propto\eta$.  The strongly correlated phases are a self-organized superradiant (SSF), a self-organized Mott insulator (SMI), a self-organized dimerized superfluid (SDSF), and a self-organized second-order superfluid (2-SSF) phase. 
	The orange line delineates superfluid phases and globally Mott insulating phases, while the green line marks the hierarchical self-organization to dimerized phases.  Pronounced sensitivity to the ramping protocol is seen in the hatched dark green region. At  higher $A$, the system is dynamically unstable to the formation of quasiperiodic attractors (QA) followed by chaos.
	The QAs only exist in the region represented by the thick gray line, while the thin dashed section represents a direct transition to the chaos. The dimensionless detuning $\delta$ and the potential strength $A$ are normalized with respect to $NU_0$ and $\sqrt{\omega_R}$, respectively.
	(c) Sketch of the SSF, SMI, SDSF, and 2-SSF phases. 
	}
	\label{fig:diagram}
\end{figure}

\begin{figure*}[t]
	\centering
	\includegraphics[width=\textwidth]{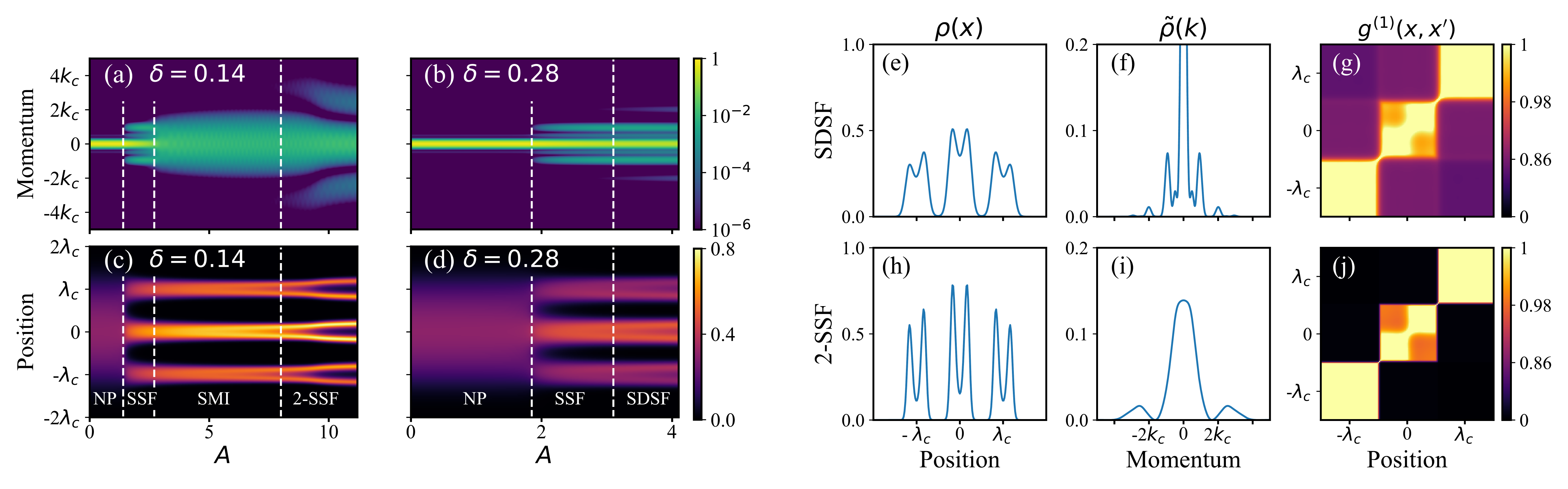}
	\caption{(a-d) The momentum  and  position space density distributions $\tilde{\rho}(k)$  and $\rho(x)$ as a function of pump rate $A\propto\eta$ at two detunings $\delta=0.14$  and $\delta=0.28$. At lower detuning, $\delta=0.14$, the system starts from the normal phase and then enters the SSF phase at $A=1.4$, the SMI phase at $A=2.7$, and the 2-SSF phase sequentially. At higher detuning $\delta=0.28$, the system starts from the normal phase and then enters the SSF phase at $A=1.9$, and the SDSF phase sequentially. The dotted lines are guides to the eye. 
	(e-j) The position  and momentum space density distributions and the Glauber one-body correlation function of  an SDSF state (first row) and a 2-SSF state (second row). In panels (g,j), the color code follows the function $-\ln(1-g^{(1)})$. The  double-well splitting  is seen in the central lattice site but not in the other two lattice sites, because only $M=4$ orbitals are used in the numerical simulations~\cite{supmat}. }
	\label{fig:SSO}
\end{figure*}

Prior theoretical work highlighted the ability of blue-detuned cavity-atom system to stabilize limit cycles and chaos~\cite{piazza15,kessler19}. Though the predicted limit cycles were not seen in the first experimental study of this regime, interesting dynamical instabilities was reported~\cite{zupancic19}.  Motivated by this study, we go beyond Refs.~\cite{piazza15,kessler19} and explore the atomic correlations and dynamical instabilities in a realistic harmonically trapped cavity-BEC system [Fig.~\ref{fig:diagram}(a)], and map the rich phase diagram in Fig.~\ref{fig:diagram}(b). On the pathway to chaos, we reveal an unexpected hierarchical deformation of the optical lattice into a double-well lattice, which generates new correlated phases of the atoms. 
In the dynamically unstable regime, we also observe that quasiperiodicity dominates instead of strict periodicity, which is compatible to experiments. Our proposed phase diagram and methodology is relevant for different experimental realizations like in Refs.~\cite{baumann10}, \cite{klinder15}, and \cite{kroeze18}.

\emph{Model and method} -  We consider a cavity-BEC system with $N$ bosonic atoms of mass $m$.  From a computational perspective, since the physics of the system does not qualitatively depend on the dimensionality, we  study a one-dimensional version of the model and later discuss the validity of the results obtained for the two-dimensional system. 
In the rotating frame of the pump laser, the system is described by the following coupled equations of motion for the cavity expectation value $\alpha$ and the atomic field operators  $\hat{\Psi}^{(\dagger)}(x)$~\cite{maschler08,nagy08}:
\begin{subequations}
	\label{eq:hamiltonian}
	\begin{eqnarray}
	i\partial_t\hat{\Psi}(x)
	&=&\left[-\frac{\hbar\partial_x^2}{2m}+\frac{g}{\hbar}\hat{\Psi}^\dagger(x)\hat{\Psi}(x)+\frac{1}{\hbar}V_\text{trap}(x)\right. \label{eom_phi} \\
	&&+\left. U_0\cos^2(k_cx)|\alpha|^2+\eta \cos(k_cx)(\alpha+\alpha^*) \right] \hat{\Psi}(x). \nonumber \\ 
	\partial_t\alpha &=&(i\Delta_c-iNU_0B-\kappa)\alpha-i\eta N\theta. \label{eom_alpha}
	\end{eqnarray}
\end{subequations}
Here, $k_c$ is the wave vector of the cavity field and corresponds to the recoil frequency $\omega_R\equiv\hbar k_c^2/2m$. $g$ is the  weak interatomic interaction, $U_0$ the atomic single photon light shift, $\eta$ the effective pump rate, $\Delta_c$ the cavity detuning, and $\kappa$ the cavity dissipation rate. The blue detuning of atoms and cavity is reflected in $U_0$ and $\Delta_c$ being positive, respectively.
The atoms are confined by a harmonic trap $V_\text{trap}(x) = \frac{m}{2}\omega_x^2x^2$. 
Since we are interested in regimes far from the normal--superradiant boundary, the cavity is in a coherent state with negligible fluctuations~\cite{nagy10,nagy11,brennecke13,lin19}, and can thus be represented by a complex number $\alpha$.
The variables $\theta=\int dx \rho(x)\cos(k_cx)$ and $B =\int dx \rho(x)\cos^2(k_cx)$ in Eq.\eqref{eom_alpha} are the order parameters associated with superradiance, where $\rho(x)=\langle \hat{\Psi}^\dagger(x)\hat{\Psi}(x)\rangle/N$ is the position space density distribution.

The main characteristic features of the system are direct results of the cavity-induced potential. For our analytical considerations, we neglect the atomic interactions, atomic correlations, and the harmonic trap, and adiabatically eliminate the cavity field 
via setting $\partial_t\alpha =0$. We find that the atoms are effectively subject to the potential~\cite{nagy08} 
\begin{eqnarray}\label{eq:hamil_steady}
V_\mathrm{cav}(x) = A^2\hbar\omega_R[2(\delta-B)\theta \cos(k_cx)+\theta^2 \cos^2(k_cx)],
\end{eqnarray}
with  $A= \eta N \sqrt{U_0}/\sqrt{[(\Delta_c-NU_0B)^2+\kappa^2]\omega_R}$ the dimensionless overall effective potential strength and $\delta=\Delta_c/NU_0$  the dimensionless cavity detuning. The cavity dynamically creates an optical lattice potential comprising two sinusoidal terms, $\cos^2(k_cx)$ and $\cos(k_cx)$, whose amplitudes are determined by the instantaneous atomic state via $\theta$ and $B$. 

With blue-detuned atoms, $U_0>0$, 
self-organization takes place in both a red-detuned cavity $\delta<0$ and a blue-detuned cavity $0<\delta<1/2$~\cite{nagy08,baumann10,piazza13}.  
In the former case, the $\cos(k_cx)$ term dominates and the atoms are localized at the lattice sites $x_n=n\pi/k_c$ with all $n$ either even or odd. However, in the latter case, the two terms in $V_\mathrm{cav}$ can be equally significant,
forming a local double well at each site. This double-well lattice can realize the Su-Schrieffer-Heeger model in a cavity-fermion system~\cite{mivehvar17}. This analysis provides the first glimpse of intrinsically different physics in the blue-detuned region.

Most phenomena in the blue-detuned cavity-BEC system, including the atomic self-organization, the double-well lattice, and the dynamical instabilities, can be revealed by the evolution in the $(B,\theta)$ phase space in discretized time. The evolution can be found by noticing that the instantaneous potential Eq.~\eqref{eq:hamil_steady} is controlled by the two order parameters from the past step, $(B_{t},\theta_{t})$, and subsequently determines the quantum state and thus the parameters in the next step, $(B_{t+1},\theta_{t+1})$. Using a Gaussian ansatz for the quantum states, the two parameters are found to evolve following
\begin{subequations}\label{eq:flow}
	\begin{eqnarray}
	\theta_{t+1}  = e^{-1/4\Omega_t}\chi_t,\quad
	B_{t+1} = \frac{1}{2} + \frac{1}{2}e^{-1/\Omega_t}(2\chi_t^2-1)
	\end{eqnarray}
	with
	\begin{eqnarray}
	\Omega_t &=& 
	\begin{cases}
	A\sqrt{|\theta_t|(B_t-\delta-|\theta_t|)/2},\, &B_t-\delta\ge |\theta_t| \\
	A\sqrt{[\theta_t^2 - (\delta-B_t)^2]/2},\, &B_t-\delta< |\theta_t|
	\end{cases} \\
	\chi_t &=& 
	\begin{cases}
	\text{sgn}(\theta_t) ,\, &B_t-\delta\ge |\theta_t| \\
	(B_t-\delta)/\theta_t, \, &B_t-\delta< |\theta_t|.
	\end{cases} 
	\end{eqnarray}
\end{subequations}

Beyond the mean-field limit, the combination of the double-well optical lattice and weak atomic interactions  results in hierarchical transitions to a series of correlated phases of matter.
We investigate the full phase diagram described by Eq.~\eqref{eq:hamiltonian} using the multi-configurational time-dependent Hartree method for indistinguishable particles (MCTDH-X)~\cite{axel16,alon08,fasshauer16,ultracold,lin20,axel20}. The simulated  $N=50$ atoms are initialized in a Thomas-Fermi-like state,
and the many-body state of the atoms coupled to the cavity  is propagated in real time. The pump rate $\eta$ is linearly ramped up at fixed detunings to reach its desired value. The simulation parameters correspond to those realized experimentally in Ref.~\cite{baumann10} and are given in detail in the supplementary material~\cite{supmat}.

\begin{figure*}[t]
	\centering
	\includegraphics[width=\textwidth]{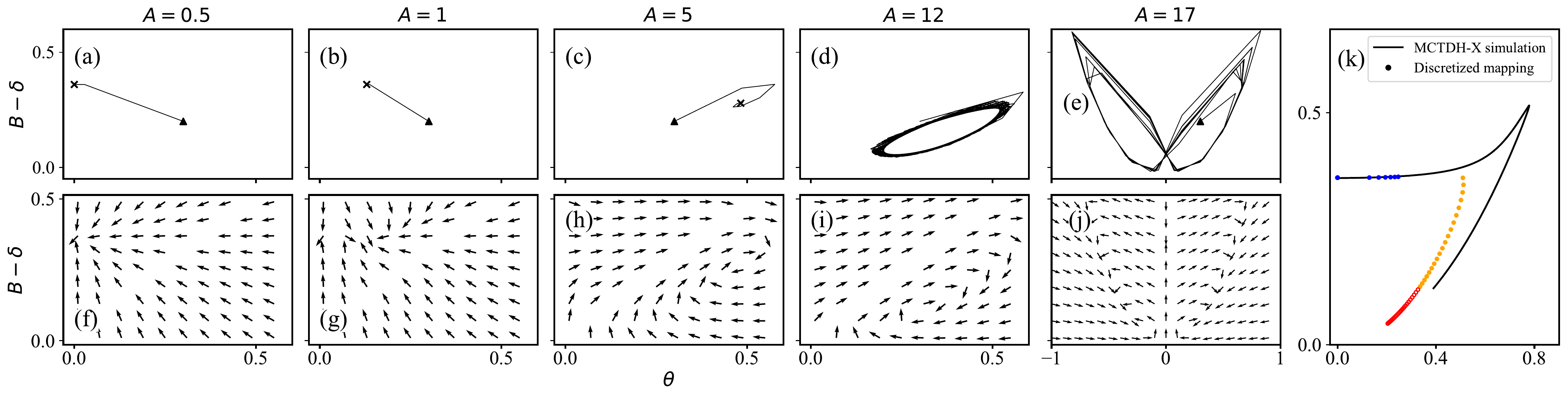}
	\caption{
		(a-e) The trajectory of the parameter pair $(B,\theta)$ on phase space evolving according to Eq.~\eqref{eq:flow}, with randomly chosen initial values marked by triangles and fixed points marked by crosses. (f-i) The corresponding flow vector plots, with normalized arrows showing only the flow directions. In all panels, we choose $\delta=0.14$. As $A$ increases, the system is in (a,f) a normal phase at $A=0.5$, (b,g) a single-well lattice at $A=1$, (c,h) a double-well lattice at $A=5$, (d,i) an attractor phase at $A=12$, and (e,j) chaos at $A=17$.
		(k) The trajectory of the fixed points as $A$ changes. The solid blue and orange points are stable fixed points for single-well and double-well lattices, respectively. The red empty points show the unstable fixed points. A blue line is superposed to show the trajectory in MCTDH-X simulation in dynamically stable phases. 
		}
	\label{fig:flow}
\end{figure*}

The phase diagram is extracted from the observables: $\theta$, $B$,  the position space density distribution $\rho(x)$, the momentum space density distribution $\tilde{\rho}(k) = \langle \hat{\Psi}^\dagger(k)\hat{\Psi}(k)\rangle/N$, and the Glauber one-body correlation function $g^{(1)}(x,x')=\langle \hat{\Psi}^\dagger(x)\hat{\Psi}(x')\rangle/\sqrt{N^2\rho(x)\rho(x')}$~\cite{glauber63,sakmann08}, whose behaviors are summarized in Fig.~\ref{fig:SSO}.
The simulation results from MCTDH-X will be compared to the analytical results from Eq.~\eqref{eq:flow} [see Fig.~\ref{fig:flow}].

\emph{Results} -  Our results for the  blue-detuned cavity-BEC are summarized in the phase diagram [Fig.~\ref{fig:diagram}(b)] along with a schematic
representation of the different phases [Fig.~\ref{fig:diagram}(c)].
The system self-organizes at a critical pump rate $\eta_c$ roughly consistent with the mean-field prediction $A(\eta_c) = 1/\sqrt{1-2\delta}$~\cite{nagy08}.   
We plot ${\rho}(x)$ and $\tilde{\rho}(k)$ at two representative detunings in Fig.~\ref{fig:SSO}(a)-(d). 
The \emph{self-organized superfluid} (SSF) phase is characterized 
by a continuous  density distribution $\rho(x)$  with pronounced peaks  at the  sites of  the emergent lattice  with spacing $\lambda_c=2\pi/k_c$.
The corresponding $\tilde{\rho}(k)$, measurable by time-of-flight experiments, is characterized by a principal peak at the center $k=0$  straddled by two satellite peaks at $k=\pm k_c$ stemming  from the superfluid  correlations between the atoms at different lattice sites~\cite{klinder15,axel17,lin19}.
At lower detunings $\delta<0.2$ and larger pump rate, the system transitions from the superfluid into a \emph{self-organized Mott insulator} (SMI) phase. 
This phase is characterized by the disappearance of the peaks at $k=\pm k_c$ and the broadening of the central peak at $k=0$ in $\tilde{\rho}(k)$  [Fig.~\ref{fig:diagram}(c)]~\cite{klinder15,greiner02,wessel04,kato08,lin19}. The superfluid and Mott insulating phases are analogue to the ones in a standard Bose-Hubbard model~\cite{lin19}.
In the $(B,\theta)$ phase space, these two phases with single-well lattices are characterized by stable fixed points with $B-\delta>|\theta|$ [Fig.~\ref{fig:flow}(b,g)]. Such phases always appear first as $A$ passes a critical value and the system leaves the normal phase $(B=1/2,\theta=0)$ [Fig.~\ref{fig:flow}(a,f)].

As the pump rate increases further, the fixed point in the phase space moves. As $B-\delta$ becomes smaller than $|\theta|$ [Fig.~\ref{fig:flow}(c,h)], local double-well potentials are formed at lattice sites according to Eq.~\eqref{eq:hamil_steady}. This is unique to blue-detuned systems.
Depending on the  degree of correlations between the atoms at different sites, we obtain either a {\it self-organized dimerized superfluid} (SDSF)  where  global superfluid correlations persist across the double-well dimers,  or a {\it self-organized second-order superfluid} (2-SSF) phase where superfluid correlations exist only within each double-well dimer. 
The signatures of these two states lie in the distributions  ${\rho}(x)$ and  $\tilde{\rho}(k)$  and the correlation function  $g^{(1)}(x,x')$ as shown in Figs.~\ref{fig:SSO}(e)-(j).  The double-well optical lattice is confirmed by the two-humped density distribution in $\rho(x)$ at each lattice site [Fig.~\ref{fig:SSO}(e,h)], and the concomitant reduction of the one-body correlation from unity within one lattice site [Fig.~\ref{fig:SSO}(g,j)].
Within each double-well dimer, local superfluidity exists and manifests itself as two peaks in $\tilde{\rho}(k)$ appearing at $k=\pm k^*$ [Fig.~\ref{fig:SSO}(f,i)], where
\begin{eqnarray}
k^*=\frac{\pi k_c}{\arccos[(B-\delta)/|\theta|]}
\end{eqnarray} 
corresponds to the distance between the minima of the double-well potential.
As the pump rate increases, $k^* $ approaches $2k_c$ from above and the peak height increases as the double well becomes deeper. 

In the SDSF (2-SSF) phase, global superfluid correlations between different pairs of double wells is present (absent). This corresponds to the presence (absence) of the peaks at $k_c$ in $\tilde{\rho}(k)$ [Fig.~\ref{fig:SSO}(f,i)], and a finite (vanishing) correlation in $g^{(1)}$ between different lattice sites [Fig.~\ref{fig:SSO}(g,j)].
In a 2-SSF state, superfluidity has a completely different length scale from the SSF and SDSF states, since coherence exists only locally within each double-well dimer.  Although superfluidity usually refers to long-range coherence, it can also be used to describe coherence within a double well~\cite{jaaskelainen05}.
These two new phases realize a variant of the Bose-Hubbard model with degenerate double-well lattices  with Hamiltonian
\begin{eqnarray}
\hat{H}_\text{BH} &=& - \sum_i (t_1\hat{c}^\dagger_{i,\mathrm{L}} \hat{c}_{i,\mathrm{R}}+t_2\hat{c}^\dagger_{i,\mathrm{R}} \hat{c}_{i+1,\mathrm{L}}+\mathrm{H.c.})\nonumber\\
&&+ \sum_{i,\sigma=\mathrm{L},\mathrm{R}}\left[\frac{U}{2}(\hat{c}^\dagger_{i,\sigma}\hat{c}_{i,\sigma})^2
+\mu_i \hat{c}^\dagger_{i,\sigma}\hat{c}_{i,\sigma}\right],
\end{eqnarray}
where $L,R$ denote the subsites~\cite{wagner12}. 

The single-well optical lattice smoothly deforms to the double-well lattice and it is hard to numerically establish if the  system transitions or crossovers during the dimerization ~\cite{supmat}.  Nonetheless, clear hysteretic behavior is seen across the boundary between SDSF and 2-SSF phases, shown by the hatched dark green region in Fig.~\ref{fig:diagram}(b). This implies a first-order transition between the globally superfluid and the globally Mott insulating phases.

\begin{figure}[t]
	\centering
	\includegraphics[width=\columnwidth]{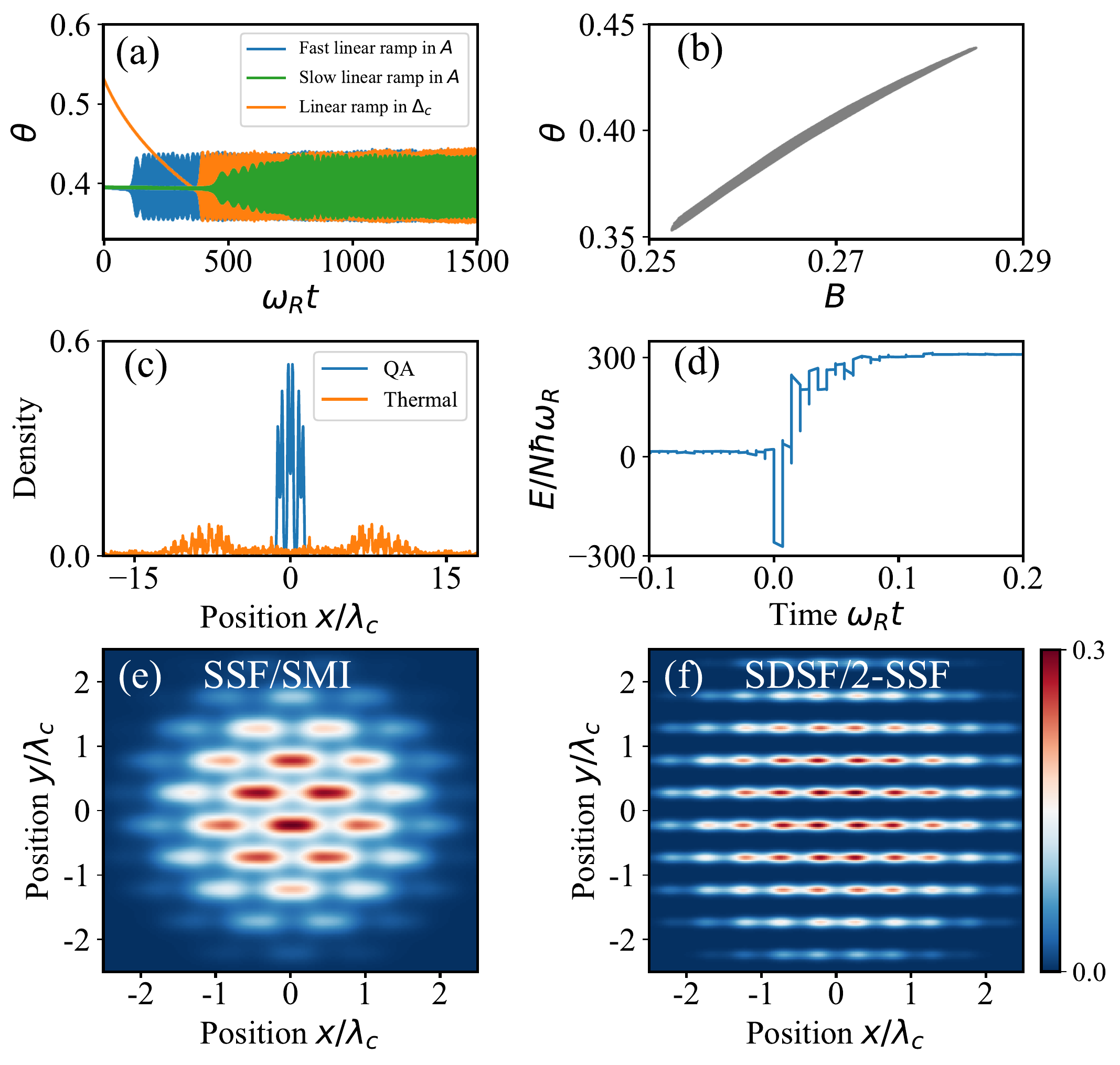}
	\caption{ 
		(a) The  order parameter $\theta$ as a function of time for three different ramping protocols from the stable region into the quasiperiodic attractor (QA) phase ($\delta=0.14$, $A=11.8$).  
		In the blue and green trajectories, the detuning is fixed at $\delta=0.14$ and the pump rate $A\propto\eta$ is ramped up linearly at rates of (blue) $dA/dt = 4\times10^{-4}\omega_R$ and (green) $dA/dt = 8\times10^{-5}\omega_R$, respectively. In the orange trajectory, the pump rate is fixed at $A=11.8$ and the detuning is ramped up linearly at a rate of $d\delta/dt = 1.5\times10^{-4}\omega_R$.
		(b) For all three  cases, the system converges to the same attractor in the $(B,\theta)$ phase space.
		(c) Density distribution $\rho(x)$ of (blue) a QA state and (orange) a thermalized state. (d)  Evolution of the system energy as the system becomes thermalized. The reference time $t=0$ is set roughly when the thermalization starts.
		(e,f) The position space density distribution $\rho(x,y)$ of (e) a single-well lattice state and (f) a double-well lattice  state in a  two-dimensional cavity-BEC system as described by Eq.~\eqref{eq:2D_pot}.
	}
	\label{fig:SA_2D}
\end{figure}

At higher pump rates, the fixed points of Eq.~\eqref{eq:flow} become unstable through a Hopf bifurcation~\cite{strogatz,andronov71} as attested by the trajectory and flow in Fig.~\ref{fig:flow}(d,i), and the appearance of dynamical instabilities in the cavity-BEC system. Such instabilities preclude Mott insulation within a double well [Fig.~\ref{fig:diagram}(b)]. 
In the presence of the atomic interactions and the harmonic trap, the limit cycles predicted in a non-interacting, trapless system~\cite{piazza15,kessler19} are reduced to \emph{quasiperiodic attractors} (QAs). 
Similar to the limit cycles, the QAs are shown to be robust against ramping protocols in Figs.~\ref{fig:SA_2D}(a,b): the trajectories are always confined in the same region in the $(B,\theta)$ phase space [Fig.~\ref{fig:SA_2D}(b)], and they exhibit roughly the same amplitude and frequency profile of oscillations [Fig.~\ref{fig:SA_2D}(a)]. In contrast to limit cycles, QAs  are highly sensitive to  initial conditions, reflecting their connection to chaos. Nevertheless, the double-well configuration of the atomic density is well preserved in a QA state [Fig.~\ref{fig:SA_2D}(c)].

As the pump rate further increases, the discretized-time model predicts $\theta$ and thereby the optical lattice will repeatedly vanish transiently [Fig.~\ref{fig:flow}(e,j)]. In the cavity-BEC system, the atoms become loosely confined and higher momentum modes are easily excited, leading to a rapid increase in system energy and resulting in full chaos and thermalization [Fig.~\ref{fig:SA_2D}(d)]. In this thermalized regime, the atomic density distribution eventually becomes completely fluctuative and the system becomes fully chaotic [Fig.~\ref{fig:SA_2D}(c)]. Tightly trapped systems are more prone to thermalization~\cite{supmat}.


 \emph{ Discussion and extension to 2D systems} - 
In terms of the renormalized parameters $\delta$ and $A$ [cf. Eq.~\eqref{eq:hamil_steady}], the phase diagram is rather general and relevant for multiple experimental setups \cite{baumann10,klinder152,kroeze18}. 
At the mean-field level, we have seen that the discretized mapping Eq.~\eqref{eq:flow} qualitatively predicts various phenomena observed in simulations. In addition, its fixed points also roughly track the simulated system trajectory in phase space [Fig.~\ref{fig:flow}(k)].

We now discuss the dependence of the phase diagram on the atom number $N$ and the trap. There are two kinds of  phase boundaries in Fig.~\ref{fig:diagram}(b):  mean-field and non-mean-field. The first kind (NP-SSF and dynamical stability boundaries) is governed by the cavity-induced effective potential in Eq.~\eqref{eq:hamil_steady}, which solely depends on $\delta$ and $A$. Thus, these phase boundaries, given as functions of $\delta$ and $A$, remain unchanged for any particle number and are only weakly sensitive to the trap~\cite{lin19}.
The second kind delineates the phases SSF, SMI, 2-SSF and SDSF. These transitions  are driven by Bose-Hubbard physics and are determined by the filling factor at lattice sites. As $N$ increases and the harmonic trap tightens, the filling factor increases and these phase boundaries move towards larger $A$. 
Based on these observations, we have made a computationally judicious choice of particle number and trapping frequency to effectively simulate the experimentally relevant phase diagram.


The phase diagram for the one-dimensional system can be straightforwardly extended to the experimentally relevant two-dimensional systems.  In this case, the atoms are subject to three light-induced effective potentials which we require to be:
\begin{eqnarray}\label{eq:2D_pot}
V_\text{light}(x,y) &=& \hbar\eta^2/U_0\cos^2(k_cy) + \hbar U_0|\alpha|^2\cos^2(k_cx) \nonumber\\
&&+\hbar\eta(\alpha+\alpha^*)\cos(k_cx)\sin(k_cy).
\end{eqnarray}
Compared to the one-dimensional version [Eq.~\eqref{eom_phi}], the first term  is an additional term stemming from the transverse pumping laser.
The cavity field $\alpha$ follows the equation of motion Eq.~\eqref{eom_alpha}, where the  lattice order parameter is generalized to  $\theta=\int dxdy \langle\hat{\Psi}^\dagger(x,y)\hat{\Psi}(x,y)\rangle\cos(k_cx)\sin(k_cy)/N$. 
Compared to the  effective potential  realized very recently in experiment~\cite{zupancic19}, the essential difference in the above system [Eq.~\eqref{eq:2D_pot}] is an extra phase shift of $\pi/2$  along the pump ($y$) axis in the last term.  This phase shift is necessary to observe our phase diagram and can straightforwardly be implemented in the experimental setup.

The two-dimensional system is expected to qualitatively possess the same  phase diagram  shown in  Fig.~\ref{fig:diagram}(b). All the phases should be accessible  for reasonable values of the detuning and pump rate.
In Figs.~\ref{fig:SA_2D}(e,f), we plot two representative examples of spatial density distributions based on mean-field simulations, i) the standard self-organization on a checkerboard lattice corresponding to an SSF or SMI, and ii)  self-organization on the double-well optical lattice structures. 
The latter would lead to extra peaks in momentum space at $\mathbf{k}=(\pm k^*,0)$ superposed over the underlying superfluid or Mott-insulator momentum distribution, which serves as the smoking-gun evidence of dimerization in experiment. 
The oscillatory  phase reminiscent of our quasiperiodic attractors has already been observed experimentally~\cite{zupancic19}. Despite the  phase shift between the experimental setup and the one discussed in this work, we expect the same mechanism driving the dynamical instabilities. To clearly observe the quasiperiodicity in an experiment, a loose trap along the cavity axis with $\omega_x\sim 10^{-3}\omega_R$  is suggested~\cite{supmat}. 
 
Our work illustrates the potential of blue detuning to realize exotic phases of matter and should stimulate future studies of complex cavity-cold atom platforms.

We acknowledge the financial support from the Swiss National Science Foundation (SNSF), the ETH Grants, Mr. Giulio Anderheggen, and the Austrian Science Foundation (FWF) under grant P32033. We also acknowledge the computation time on the ETH Euler cluster and the Hazel Hen Cray cluster at the HLRS Stuttgart.

\bibliographystyle{apsrev}
\bibliography{References}
 
\end{document}


\title{Supplementary Material for \\
	Pathway to chaos through hierarchical superfluidity \\ in blue-detuned cavity-BEC systems}
 
\author{Rui Lin}
\affiliation{Institute for Theoretical Physics, ETH Zurich, 8093 Zurich, Switzerland}
\author{Paolo Molignini}
\affiliation{Institute for Theoretical Physics, ETH Zurich, 8093 Zurich, Switzerland}
\affiliation{Clarendon Laboratory, University of Oxford, Parks Road, Oxford OX1 3PU, United Kingdom}
\author{Axel U. J. Lode}
\affiliation{Institute of Physics, Albert-Ludwig University of Freiburg, Hermann-Herder-Stra{\ss}e 3, 79104 Freiburg, Germany}
\author{R. Chitra}
\affiliation{Institute for Theoretical Physics, ETH Zurich, 8093 Zurich, Switzerland}

 


\maketitle

\section{Multiconfigurational time-dependent Hartree method}

In this work, the multiconfigurational time-dependent Hartree method for indistinguishable particles~\cite{axel16,alon08,fasshauer16,axel20} (encoded in the software MCTDH-X~\cite{ultracold,lin20}) is used to evolve the system's state. This algorithm is able to solve a Hamiltonian with a one-body potential $V(x)$  and a two-body interaction $W(x,x')$,
\begin{align} 
\hat{\mathcal{H}}=\int \mathrm{d}x \hat{\Psi}^\dagger(x) \left\{\frac{p^2}{2m}+V(x)\right\}\hat{\Psi}(x) +\frac{1}{2}\int \mathrm{d}x \hat{\Psi}^\dagger(x)\hat{\Psi}^\dagger(x')W(x,x')\hat{\Psi}(x)\hat{\Psi}(x').
\end{align}
MCTDH-X is based on the following ansatz for the many-body wave function
\begin{eqnarray}
|\Psi\rangle=\sum_{\mathbf{n}} C_\mathbf{n}(t)\prod^M_{k=1}\left[ \frac{(\hat{b}_k^\dagger(t))^{n_k}}{\sqrt{n_k!}}\right]|0\rangle ,
\end{eqnarray}
where $N$ is the number of atoms, $M$ is the number of single-particle wave functions (orbitals) and $\mathbf{n}=(n_1,n_2,...,n_k)$ gives the number of atoms in each orbital with the constraint $\sum_{k=1}^M n_k=N$. The creation operator of the atomic field $\hat{\Psi}^\dagger$ is composed by the creation operators of the individual orbitals $\hat{b}^\dagger_k$,
\begin{eqnarray}
b_i^\dagger(t)&=&\int \psi^*_i(x;t)\hat{\Psi}^\dagger(x;t) \mathrm{d}x \\
\hat{\Psi}^\dagger(x;t)&=&\sum_{i=1}^M b^\dagger_i(t)\psi_i(x;t), \label{eq:def_psi}
\end{eqnarray}
where $\psi_i(x;t)$ are the wave functions of the individual working orbitals.
The state is evolved by propagating the coefficients $C_\mathbf{n}(t)$ and the working orbitals $\psi_i(x;t)$ using the time-dependent variational principle~\cite{TDVM81}.

In MCTDH-X, the number of orbitals $M$ is important  to obtain  converged   results, especially of  correlation functions~\cite{alon08,axel12,fasshauer16}. With a single orbital $M=1$,  the method  reproduces the Gross-Pitaevskii mean-field limit; while with infinitely many orbitals $M\rightarrow\infty$, the method is numerically exact. To simulate a Mott insulating state, the number of required orbitals is usually the same as the number of \emph{lattice sites}~\cite{axel162}. In our case, considering that there are three lattice sites and each site splits into two sub-sites, $M=6$ orbitals are needed in total to fully capture the atomic correlations. However, due to the amount of computational time needed, the simulations are performed with $M=3$ orbitals or $M=4$ orbitals when specified. In the latter case, we are able to observe the effect of the correlation induced  dimerization on the one-body correlation function inside the central lattice site, as shown in Fig.~2(g,j) in the main text. We expect that with $M=6$ orbitals, reduction of correlation from unity can also be seen in the other two lattice sites.

 \section{System Parameters}

The parameters used in this work are inspired by the experimental setup in Ref.~\cite{baumann10}. We simulate a total number of $N=50$ $^{87}$Rb atoms with mass $m=1.44\times10^{-25}$kg and one-dimensional contact interaction $Ng=1.0\times10^{-17}\mathrm{eV}\cdot\mathrm{m}$.
The wavelength of the laser pump and the cavity field is chosen as $\lambda_c=784$nm, and therefore the recoil frequency is $\omega_R=2\pi\times3734$Hz. 
The collective photon light shift is given by $NU_0=28.9 \omega_R=2\pi\times 108$kHz, and the cavity detuning takes the values between $\Delta_c=0$ and $\Delta_c=NU_0/2 =2\pi\times54$kHz. The maximum pump rate is $\eta=2\pi\times 825$kHz, this corresponds to an overall effective potential strength of $A=12$ [cf. Eq.(2) in the main text]. The cavity dissipation rate is chosen to be large enough $\kappa=700 \omega_R=2\pi\times2.60$MHz such that the system operates in the bad-cavity limit.  We note that  the numerical simulations can also be done for less dissipative cavities.
The harmonic trap confining the atoms  has a trapping frequency of $\omega_x=0.136 \omega_R=2\pi\times 504$Hz in most of the simulations, but looser traps with $\omega_x=0.068\omega_R=2\pi\times252$Hz and $\omega_x=0.0068\omega_R=2\pi\times25.2$Hz are also used to analyze the quasiperiodic attractor phase.

\section{Dynamical equations for $\theta$ and $B$ in discretized time}
The dynamical behavior of the system can already be captured by the evolution of the parameters $\theta$ and $B$.  These two parameters are chosen because they enter the effective Hamiltonian [cf. Eq.(2) in the main text] and they quantify the self-organization.
The discrete dynamical map governing the evolution of these two parameters can be derived based on the effective one-body potential, Eq.~(2) in the main text:
\begin{eqnarray}
\hat{H}^{(1)} &=& -\frac{\hbar\partial_x^2}{2m}+A^2\hbar\omega_R[2(\delta-B)\theta \cos(k_cx)+\theta^2 \cos^2(k_cx)].
\end{eqnarray}
Due to the driven-dissipative nature of the system, the atoms are subject to an effective potential which is determined by the atomic distribution itself through $\theta$ and $B$. On the other hand, once the instantaneous effective potential is known, we are able to solve the ground state of the Hamiltonian, which will give a new pair of parameters $\theta$ and $B$. In this way, we are able to obtain discretized evolution equations for these two parameters. 

We suppose that at step $t$, the parameters are given by $\theta_t$ and $B_t$.
When $B-\delta\geq|\theta|>0$, the minima of the effective potential lie at $x_n = 2n\pi/k_c$ [$x_n = (2n+1)\pi/k_c$], $n\in\mathbb{N}$, for positive (negative) $\theta$. In the vicinity of the minima, the effective potential can be expanded into a quadratic form
\begin{eqnarray}
V_t(x = x_n+\delta x) = \hbar\omega_RA^2|\theta_t|(B_t-\delta-|\theta_t|) k_c^2\delta x^2\equiv \frac{m}{2}\varpi_t^2\delta x^2.
\end{eqnarray}
The atomic density distribution at the next step can immediately be written down using the Gaussian ansatz for the ground state:
\begin{eqnarray}
\rho_{t+1}(x) \propto \sum_{n} \exp\left[-\frac{m\varpi_t (x-x_n)^2}{\hbar}\right].
\end{eqnarray}
Therefore, the parameters $\theta$ and $B$ at the next step can be calculated according to their definitions, $\theta_{t+1}=\int_{-\infty}^{\infty} \cos(k_cx)\rho_{t}(x)$ and $B_{t+1}=\int_{-\infty}^{\infty} \cos^2(k_cx)\rho_{t}(x)$, respectively. According to the aforementioned ansatz of the density distribution, these are given explicitly by
	\begin{eqnarray}
	\theta_{t+1} = \text{sgn}(\theta_{t})\exp\left[-\frac{\hbar k_c^2}{4m\varpi_t}\right],\quad 
	B_{t+1} = \frac{1}{2}+\frac{1}{2}\exp\left[-\frac{\hbar k_c^2}{m\varpi_t}\right]
	\end{eqnarray}
We note that $\varpi_t$ defined here and $\Omega_t$ from the main text are related through $\Omega^2_t = m\varpi^2_t/\hbar k_c$.

When $B-\delta\geq|\theta|<0$, the cavity-induced potential forms a double-well lattice, but the evolution equations for $\theta$ and $B$ can be found in a similar manner as above. The results are summarized in Eq.~(3) of the main text.

\section{ Crossover  or transition to  SDSF and 2-SSF phases }\label{sec:dimer_op}

\begin{figure}[h]
	\centering
	\includegraphics[width=0.6\columnwidth]{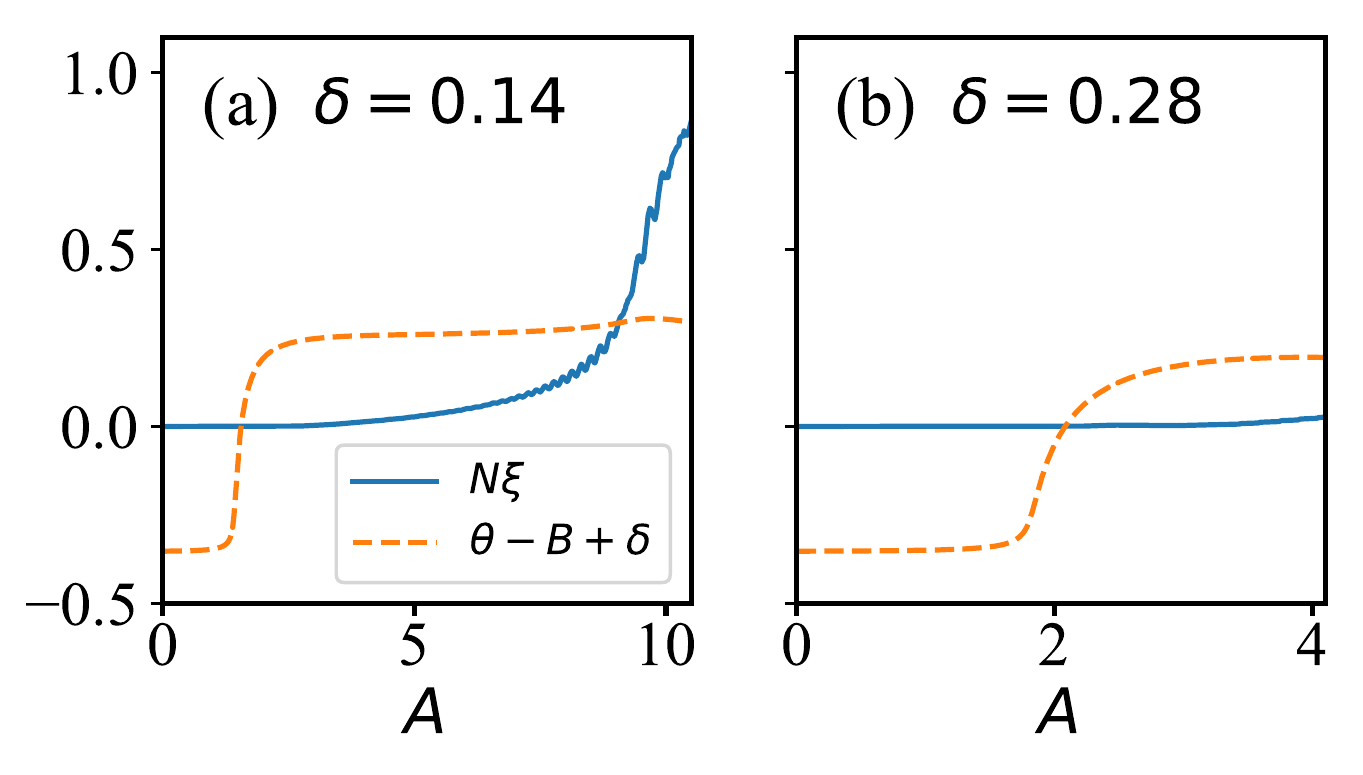} 
	\caption{The dimerization order parameter $\xi$ [Eq.~\eqref{eq:dimer_op}] and $\theta-B+\delta$ as functions of pump rate at different detunings (a) $\delta=0.14$ and (b) $\delta=0.28$. The dimerization order parameter $\xi$ is multiplied by a factor of $N=50$ such that it is in the same order of magnitude as $\theta-B+\delta$. The dimensionless potential strength $A$ is given with respect to $\sqrt{\omega_R}$.
	}
	\label{fig:dimer_OP}
\end{figure}

To investigate if the spontaneous deformation to the double-well lattice signifies a transition or a crossover, we  choose
the height of the  peak at $k=k^*$ in momentum space as the  relevant order parameter,
\begin{eqnarray}\label{eq:dimer_op}
\xi = \begin{cases}
0, & B-\delta \ge |\theta|\\
\tilde{\rho}(k^*), & B-\delta < |\theta|,
\end{cases}
\end{eqnarray}
where  $k^* = \pi k_c/\arccos[(B-\delta)/|\theta|]$.
This  ad hoc  order parameter $\xi$ as a function of $A$ at two different detunings $\delta=0.14$ and $\delta=0.28$ is shown in Fig.~\ref{fig:dimer_OP}. As discussed in the main text,  for  $\delta=0.14$  the system enters the normal phase (NP), the self-organized superfluid (SSF) phase, the self-organized Mott insulator (SMI) phase and the self-organized second-order superfluid phase (2-SSF) phase sequentially. In the latter case, the system enters the NP, SSF phase and self-organized dimerized superfluid (SDSF) phase sequentially.

In both cases, $\xi$ gains a tiny finite value as soon as $\theta>B-\delta$, as the effective potential [Eq.~(2) in the main text] at each lattice site evolves from a single well into a double well. However, at $\delta=0.14$, $\xi$ increases slowly at first in the SSF and SMI phases and then it increases rapidly after entering the 2-SSF phase. The  onset of the rapid increase  ($A\approx9$)  roughly traces the boundary between the SMI and 2-SSF phases. On the contrary, at $\delta=0.28$, the increase in $\xi$ is almost negligible. It is more likely that the system undergoes a crossover between the SSF and SDSF phases and there is no clear boundary between these two phases.

Tracing through the green line indicating the dimerization transition in the phase diagram [Fig.~1(b) in the main text], we observe two different transition behaviors depending on the existence of the global superfluidity. When the detuning is large and the system is globally superfluid, a crossover is likely to take place; while when the detuning is small and the system is globally Mott insulating, a second order transition is likely to take place. A firm conclusion can be drawn only after more detailed examination, including the dependence of the transition behaviors on atom number and orbital number.

\section{Hysteresis in atomic correlations between the two dimerized phases}

\begin{figure}[h]
	\centering
	\includegraphics[width=\textwidth]{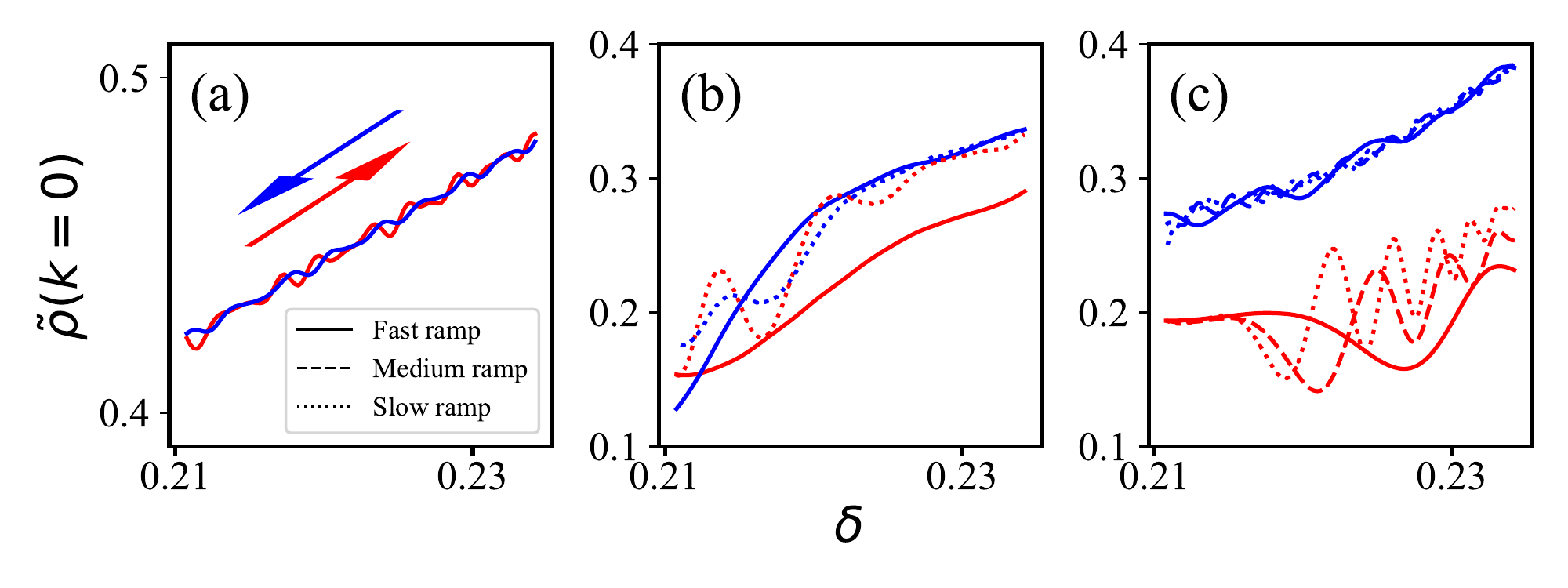}
	\caption{The hysteretic behaviors across the SDSF - 2SSF boundary. The cavity detuning is ramped up (red lines) and down (blue lines) between $\delta=0.211$ ($\Delta_c=45$) and $\delta=0.234$ ($\Delta_c=50$) at different pump rates (a) $A = 2.9$, ramping within SSF, (b) $A= 4.4$, and (c) $A= 5.9$. The ramping is performed at different rates, where the fastest ramp is performed within a time interval of $t=13.6/\omega_R$ (solid lines), the medium ramp within $t=40.7/\omega_R$ (dashed lines), and the slowest ramp within $t=67.8/\omega_R$ (dotted lines). The dimensionless cavity detuning $\delta$ is given with respect to $NU_0$.}
	\label{fig:hysteresis}
\end{figure}

We now analyze the hatched dark green region separating the self-organized dimerized superfluid (SDSF) phase and the self-organized second-order superfluid (2-SSF) phase in Fig.~1(b) in the main text.
A systematic analysis  shows that  the correlations of the system  hardly converge close to the boundary between the SDSF  and 2-SSF phases.  
The height of the central peak in the momentum space density distribution $\tilde{\rho}(k=0)$ has been shown in the literature to serve as a simple and useful indicator of the correlations between the atoms~\cite{kato08,wessel04,lin19}.
In Fig.~\ref{fig:hysteresis}, we fix the pump rate at three different values and ramp the cavity detuning back and forth across the SDSF - 2SSF boundary at different ramping rates. 
At small pump rate $A = 2.9$ [Fig.~\ref{fig:hysteresis}(a)], the system is still in the SSF phase and no hysteresis exists even for the fastest ramp. 
At higher pump rate $A= 4.4$ [Fig.~\ref{fig:hysteresis}(b)], the system enters the SDSF phase and the 2SSF phase at high and low detunings, respectively. 
A hysteresis in $\tilde{\rho}(k=0)$ can be clearly seen in a fast ramping. 
However, the hysteresis area becomes much smaller as the ramping slows down. 
At an even higher pump rate $A = 5.9$ [Fig.~\ref{fig:hysteresis}(c)] where the system is approaching the thermalized phase, the hysteresis is prominent, and the hysteresis area does not vanish even for the slowest ramping.
Our results indicate that the SDSF--2-SSF transition could be a first-order transition.

\section{Sensitivity of the quasiperiodic attractor to  the harmonic trap and the atomic fluctuations}\label{sec:sensitivity}

\begin{figure}[h]
	\centering
	\includegraphics[width=0.8\columnwidth]{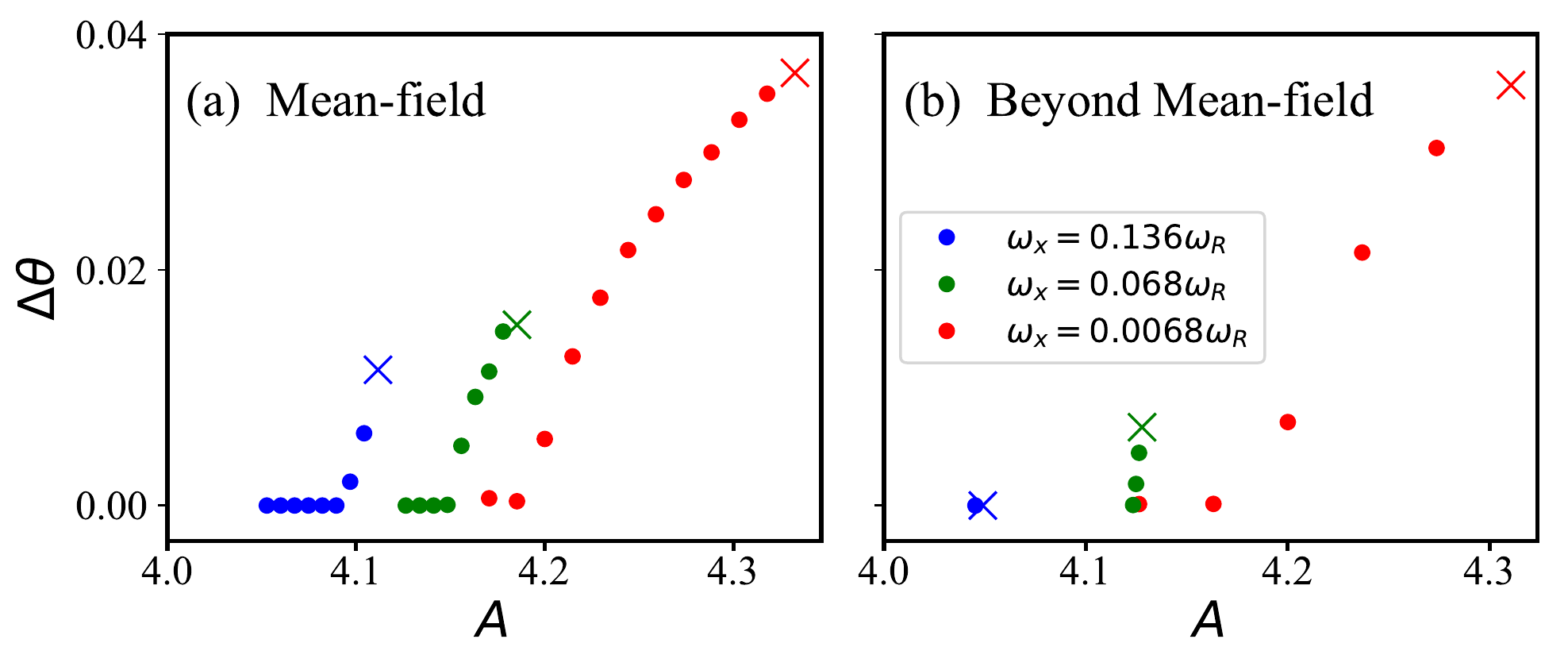}
	\caption{ 
		Measurement of the variance of the order parameter $\Delta\theta$ [Eq.~\eqref{eq:delta_theta}] over time as the system is propagated at different pump rate at $\delta=0.4$. The propagation is done with different trapping frequencies ($\omega_x/\omega_R=0.136$, $0.068$, and $0.0068$) both (a) in the mean-field limit and (b) beyond. 
		The crosses mark the largest value of pump rate where the system is not thermalized. The dimensionless potential strength $A$ is given with respect to $\sqrt{\omega_R}$.
	}
	\label{fig:stable_unstable}
\end{figure}

The presence of trapping and atomic interactions have large impact on the transition to chaos. The quasiperiodic attractor (QA) is highly susceptible to the harmonic trap and atomic fluctuations.
To quantify the dynamical instability in the system, we propagate the system  to the  desired  pump rate  using different ramp-up times.   We then compute  the magnitude of the system's oscillation through the variance of $\theta$, i.e., 
\begin{eqnarray}\label{eq:delta_theta}
\Delta\theta=\sqrt{\overline{\theta^2}-\overline{\theta}^2},
\end{eqnarray}
where the bar means average in time.  For comparison, these propagations are performed with different values of the harmonic trapping frequency, both in the mean-field limit [Fig.~\ref{fig:stable_unstable}(a)] and  beyond mean-field [Fig.~\ref{fig:stable_unstable}(b)].
We find that a  looser trap shifts the boundary between the stable and unstable region towards higher pump rates and also widens the QA region. 
In the mean-field limit, as the trap loosens, the width of the QA region ($\Delta A$) becomes wider. 
We compare our results with the mean-field results in Ref.~\cite{piazza15}, where $NU_0=12.1\omega_R$, $\kappa=10\omega_R$ and the trap is absent. At the same detuning $\delta=0.4$, i.e. $\Delta_c=\delta NU_0=4.84\omega_R$, the limit cycle region has been predicted to take place between roughly between $\eta\sqrt{N}=8.5\omega_R$ and $11\omega_R$. This corresponds to roughly $A=3.0$ and 3.8.
The QA region shrinks even further as we turn on the atomic fluctuations. 
Particularly, with the tight trapping, $\omega_x/\omega_R=0.136$, the QA region between the SDSF phase and the thermalized region vanishes completely.
In conclusion, the quasiperiodic attractors are extremely susceptible to the atomic fluctuations and the harmonic trap frequency. A tight trap can drive the quasiperiodic attractors into chaos. In experiments, a loose trap with $\omega_x\sim 10^{-3}\omega_R$ is recommended for the observation of the quasiperiodic attractors.

\section{System trajectory in higher-dimensional phase space and the process of thermalization}
\label{sec:thermal}
\begin{figure}[h]
	\centering
	\includegraphics[width=0.7\textwidth]{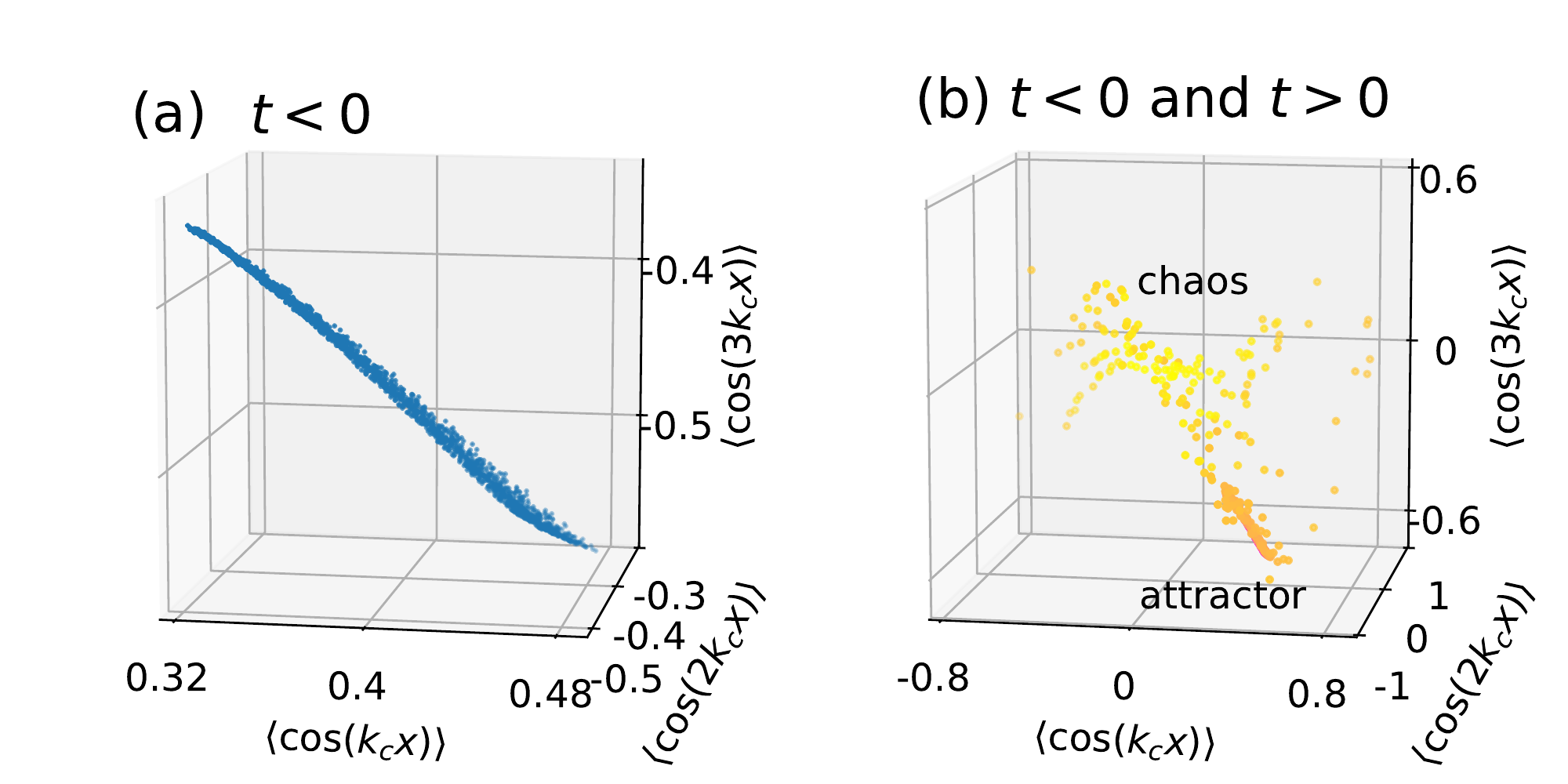}
	\caption{The trajectory of the system projected on the basis $\{\langle\cos(k_cx)\rangle, \langle\cos(2k_cx)\rangle, \langle\cos(3k_cx)\rangle\}$ (a) before thermalization and (b) both before and after thermalization. In panel (b), the color scheme shows the time. The trajectory evolves starting from the orange region and ending in the yellow region. The simulation is performed with $M=3$ at detuning $\delta=0.18$.}
	\label{fig:therm}
\end{figure}



Apart from the density distribution and the system energy, the quasiperiodic attractor phase and the thermalized phase also differ vastly in their trajectory in the phase space. Particularly, such difference is more pronounced in the high-dimensional phase space.
We thus expand the density distribution in the basis $\{\langle\cos(k_cx)\rangle=\theta, \langle\cos(2k_cx)\rangle=2B-1, \langle\cos(3k_cx)\rangle, \langle\cos(4k_cx)\rangle,\dots\}$. The projections of the trajectory onto the first three basis vectors $\langle\cos(k_cx)\rangle$, $\langle\cos(2k_cx)\rangle$, and $\langle\cos(3k_cx)\rangle$ in the quasiperiodic attractor regime and the chaotic regime are shown in Fig.~\ref{fig:therm}. 

Fig.~\ref{fig:therm}(a) shows the system trajectory before thermalization. The oscillation along the third direction $\langle\cos(3k_cx)\rangle$ has a significant dependence on the oscillation along the first two directions. To analyze such a dependence, we perform principle component analysis (PCA) on the projection on the first six basis vectors $\langle\cos(k_cx)\rangle,\dots,\langle\cos(6k_cx)\rangle$. The eigenvalues of the six components are $\lambda_1=0.12$, $\lambda_2=0.02$, $\lambda_3=0.01$, $\lambda_4=0.004$, $\lambda_5=7\times10^{-4}$, $\lambda_6=9\times10^{-5}$. The fast decrease in these eigenvalues suggests that the quasiperiodic attractor is essentially a ``limit tube'' in the Hilbert space.

Fig.~\ref{fig:therm}(b) shows the system trajectory both before and after thermalization. After the system becomes thermalized, the system trajectory moves to a completely different part of phase space. In this scenario, the eigenvalues of the six components in PCA analysis are $\lambda_1=0.071$, $\lambda_2=0.034$, $\lambda_3=0.032$, $\lambda_4=0.025$, $\lambda_5=0.021$, $\lambda_6=0.017$. The decrease in the eigenvalues is much slower than the previous case, suggesting a high trajectory dimension. 
The significant difference between the quasiperiodic attractor regime and the chaotic regime reassures that they belong to two dynamical phases.

\section{Bifurcation related to the dynamical instabilities}
\label{sec:analytics}

\begin{figure}[h]
	\centering
	\includegraphics[width=0.6\textwidth]{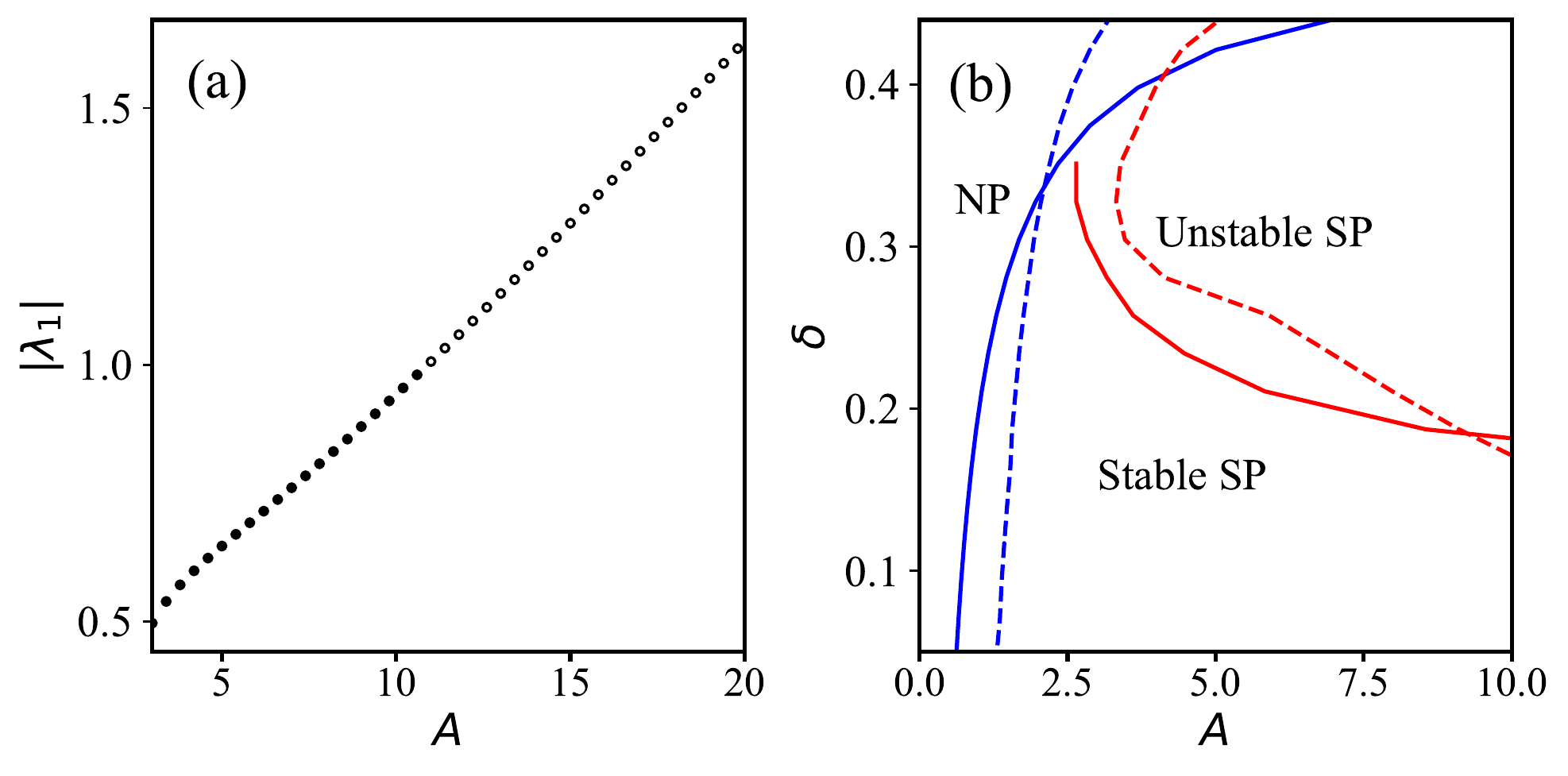}
	\caption{
		(a) The modulus of the eigenvalue of the Jacobian $|\lambda_1|$ as a function of $A$. As $A$ increases, it grows from below 1 where the fixed point is stable (solid points) to above 1 where the fixed point becomes unstable (empty points). In particular, it reaches $|\lambda_1|=1$ at roughly $A=11$, where the system undergoes a Neimark-Sacker bifurcation. The detuning takes the value of $\delta=0.14$.
		(b) The mean-field phase diagram obtained by (solid lines) the discretized time analysis, and (dashed lines) MCTDH-X simulations. The simulation results are reproduced from Fig.~1(b) of the main text. The blue lines show the boundary between the normal phase (NP) and the superradiant phase (SP), while the red lines separate the dynamically stable and unstable regions. 
		For small detunings $\delta<0.3$, the system behaviors are qualitatively the same between the discretized time analysis and the simulations. For large detunings $\delta>0.3$, the discretized time analysis predicts that the system will become dynamically unstable as soon as it enters the SP. The dimensionless detuning $\delta$ and the potential strength $A$ are normalized with respect to $NU_0$ and $\sqrt{\omega_R}$, respectively.
	}
	\label{fig:flow_eigen}
\end{figure}

The evolution of the system in discretized time and in the $(B,\theta)$ phase space is described by Eq.~(3) of the main text. It also helps for investigating the nature of the bifurcation in the cavity-BEC system. The dynamical behavior of a system can be studied by linearization in the vicinity of the fixed point, which is performed via the Jacobian
\begin{eqnarray}
\label{eq:jacobian}
J = \begin{pmatrix}
\dfrac{\partial\theta_{t+1}}{\partial\theta_{t}} & \dfrac{\partial\theta_{t+1}}{\partial B_{t}}\\
\dfrac{\partial B_{t+1}}{\partial\theta_{t}} & \dfrac{\partial B_{t+1}}{\partial B_{t}}
\end{pmatrix}.
\end{eqnarray}
The eigenvalues of the Jacobian are denoted as $\lambda_1$ and $\lambda_2$. After the formation of the double-well lattice with $\delta-B<|\theta|$, we find that the two eigenvalues are complex conjugates $\lambda_1=\lambda_2^*$, indicating a spiral fixed point. In a discretized system, a spiral fixed point goes from stable to unstable as the moduli of its eigenvalue $|\lambda_1|=|\lambda_2|$ passes through 1 from below to above. This is called a Neimark-Sacker bifurcation~\cite{neimark59,sacker64}, whose conterpart in continuous time is the Hopf bifurcation~\cite{strogatz,andronov71}.
In our model, for detuning $\delta=0.14$, the bifurcation take place at roughly $A=11$ [Fig.~\ref{fig:flow_eigen}(a)]. Such a bifurcation always takes place when $|\theta|>B-\delta$ and is thus closely related to the formation of double-well lattices.

As we recall from the main text, the normal--superradiant boundary and the dynamically stable--unstable boundary are solely driven by the cavity-induced potential and can be fully treated in the mean-field limit. This indicates that the discretized mapping should also fully capture these two boundaries. In Fig.~\ref{fig:flow_eigen}(b), we compare these two boundaries obtained from the discretized system and the simulation results of the cavity-BEC system by MCTDH-X. Indeed, the two results roughly agree with each other, especially at small detunings.

\bibliographystyle{apsrev}
\bibliography{References}